\documentclass[iop]{emulateapj}
\usepackage{color}
\usepackage{graphicx}
\usepackage{amsmath}

\shorttitle{Direct measurement of the distribution of dark matter}
\shortauthors{Cao, et al.}

\begin{document}
\title{DECi-hertz Interferometer Gravitational-wave Observatory: Forecast constraints on the cosmic curvature with LSST strong lenses}
\author{Shuo Cao\altaffilmark{1$\ast$}, Tonghua Liu\altaffilmark{2}, Marek
Biesiada\altaffilmark{3}, Yuting Liu\altaffilmark{1}, Wuzheng Guo\altaffilmark{1}, Zong-Hong
Zhu\altaffilmark{1$\dag$}}

\altaffiltext{1}{Department of Astronomy, Beijing Normal University,
100875, Beijing, China; \emph{caoshuo@bnu.edu.cn; zhuzh@bnu.edu.cn}}
\altaffiltext{2}{School of Physics and Optoelectronic, Yangtze University, Jingzhou 434023, China;}
\altaffiltext{3}{National Centre for Nuclear Research, Pasteura 7,
02-093 Warsaw, Poland}

\begin{abstract}

In this paper, we aim at using the DECi-hertz Interferometer Gravitational-wave Observatory (DECIGO), a future Japanese space gravitational-wave antenna sensitive to frequency range between LISA and ground-based detectors, to provide gravitational-wave constraints on the cosmic curvature at $z\sim 5$. In the framework of the well-known distance sum rule, the perfect redshift coverage of the standard sirens observed by DECIGO, compared with lensing observations including the source and lens from LSST, makes such cosmological-model-independent test more natural and general. Focusing on three kinds of spherically symmetric mass distributions for the lensing galaxies, we find that the cosmic curvature is expected to be constrained with the precision of $\Delta \Omega_K \sim 10^{-2}$ in the early universe ($z\sim5.0$), improving the sensitivity of ET constraints by about a factor of 10. However, in order to investigate this further, the mass density profiles of early-type galaxies should be properly taken into account. Specially, our analysis demonstrates the strong degeneracy between the spatial curvature and the lens parameters, especially the redshift evolution of power-law lens index parameter. When the extended power law mass density profile is assumed, the weakest constraint on the cosmic curvature can be obtained. Whereas, the addition of DECIGO to the combination of LSST+DECIGO does improve the constraint on the luminosity density slope and the anisotropy of the stellar velocity dispersion significantly. Therefore, our paper highlights the benefits of synergies between DECIGO and LSST in constraining new physics beyond the standard model, which could manifest itself through accurate determination of the cosmic curvature.

\end{abstract}


\maketitle

\section{Introduction} \label{introduction}

The question of whether the Universe is spatially open, flat or
closed, which can be quantitatively addressed by determining the
cosmic curvature parameter (hereafter $\Omega_k$) has been the
subject of recent intense discussions. Nowadays, a large amount of
independent observations have provided strong evidence supporting
the spatial flatness of our Universe within the current precision
\citep{Cai2016,Li2016c,Wei2017,Wang2017,Rana2017}, which is well
consistent with the predictions of different inflationary models
\citep{Ichikawa2006,Virey2008}. However, the recent Planck 2018
results, which provided the latest measurements of temperature and
polarization of the cosmic microwave background anisotropy
\citep{Planck Collaboration}, tend to favor a spatially closed
Universe over $2\sigma$ confidence level
($\Omega_k=-0.044^{+0.018}_{-0.015}$). It should be pointed out that
in several recent works \citep{Liu19} the validity of the result
might be as problematic, due to its strong dependency on the assumed
non-flat $\Lambda$CDM model. Such tension becomes one of the major
puzzles in modern cosmology. On the one hand, the constraints on
$\Omega_k$ also have important consequences for properties of dark
energy \citep{Clarkson2007,Gong2007}, considering the correlation
between the cosmic curvature and dark energy model used in fitting
different observational data, such as the Baryon acoustic
oscillation, Hubble parameter, and angular size measurement
\citep{Ryan19}. Others have argued that the derived values of the
cosmic curvature are highly dependent on the validity of the
background FLRW metric, a more fundamental cosmological assumption
which has been investigated in many recent studies with strongly
lensed SNe Ia and gravitational waves \citep{Denissenya2018,Cao19a}.
In any case, in order to better understand the curvature tension and
the nature of dark energy, it is necessary to emphasize the
importance of determining model-independent measurements of the
spatial curvature with different geometrical methods. This could
also be reason why providing and forecasting fits on $\Omega_k$ from
current and future astrophysical observations has become an
outstanding issue in modern cosmology
\citep{Cai2016,Li2016c,Wei2017,Wang2017,Rana2017}.

In this paper, we focus on the Distance Sum Rule (DSR) in the
framework of strong gravitational lensing (SGL) by early-type
galaxies \citep{Takada2015,Denissenya2018,Ooba2018}. More
specifically, the ratios of two angular diameter distance
$D_{ls}/D_s$ (the source-lens/lens distance ratio) can be directly
inferred from the observations of Einstein radii, with precise
measurements of central velocity dispersions of the lensing galaxies
\citep{Bolton08,Cao2012,Cao2015b}. Meanwhile, the distances at
redshifts $z_l$ and $z_s$ are always measured from several popular
distance indicators covering these redshifts, such as SNe Ia, Hubble
parameters \citep{Clarkson2008,Clarkson2007,Shafieloo2010,Li2016c}
and intermediate-luminosity radio quasars \citep{Cao17a,Cao17b}
acting as standard candles, cosmic chronometers and standard rulers,
respectively. However, the uncertainty of the latest $\Omega_k$
constraint was quite large due to the limited sample size of
available SGL data \citep{Xia2017,Qi2019}, focusing on 118
galactic-scale strong lensing systems from the Sloan Lens ACS Survey
(SLACS), BOSS emission-line lens survey, Lens Structure and
Dynamics, and Strong Lensing Legacy Survey \citep{Cao2015b}.
Besides, only a small fraction of the lensing data can be utilized,
due to the mismatch of redshifts between the lensing systems
(especially the background sources) and the current SNe Ia sample.
Such disadvantage of this method will be more severe in the near
future, when the source redshift of galactic-scale strong lensing
systems is expected to reach $z\sim 5$ in the forthcoming Large
Synoptic Survey Telescope (LSST) \citep{Oguri10,Vermai2019}.
Therefore, the direct luminosity distances with high redshifts would
significantly contribute to a robust measurement of the cosmic
curvature, which has been demonstrated in a recent analysis of UV
and X-ray quasars \citep{Risaliti2018}.

On the other hand, in the gravitational wave (GWs) domain, one could
use the GW signals from inspiralling and merging compact binaries to
derive luminosity distances \citep{Schutz86}. Such methodology has
been realized by Advanced LIGO and VIRGO detectors, with the
detection of different types of signals including a binary neutron
star system \citep{Abbott16,Abbott17}. Specially, the Hubble diagram
of these so-called standard sirens could be directly constructed and
applied in cosmology, with the redshifts measured from their
electromagnetic (EM) counterparts. Looking ahead, the next
generation detectors like the DECi-Hertz Interferometric
Gravitational Observatory (DECIGO), a future Japanese space GW
antenna, will extend the detection limit of Advanced LIGO and
Einstein Telescope to the earlier stage of the Universe ($z\sim 5$)
generating the detections of $\sim 10^4$ NS-NS binaries per year. In
addition, the detection of these binary systems using the
second-generation technology of space-borne DECIGO takes place in
the inspiral phase long time ($\sim 5$ years) before they enter the
LIGO frequency range, with signal-noise-ratio (SNR) much higher than
that of the current and future ground-based GW detectors. Therefore,
in this study we propose with the future standard siren sample from
DECIGO, the largest compilation of SGL data expected from LSST can
be used to infer the cosmic curvature resulting in more precise
constraints. This paper is organized as follows. In Sec. 2 and 3, we
will briefly introduce the methodology and the simulated data
(DECIGO standard sirens and LSST strong lenses) in this analysis.
The forecasted constraints on the cosmic curvature are presented in
Sec. 4. Finally, we give summaries and discussions in Sec. 5.
Throughout the paper, the flat $\Lambda$CDM is taken as the fiducial
cosmological model in the simulation, with $\Omega_m=0.315$ and
$H_0=67.4$ km/s/Mpc from the latest \textit{Planck} observations
\citep{Planck Collaboration}.

\section{Methodology}

As one of the basic assumptions in cosmology, the cosmological
principle (i.e., the Universe is homogeneous and isotropic at large
scales) has been widely applied in different cosmological studies.
Now the FLRW metric is introduced to describe the space-time of the
Universe (where the speed of light $c=1$)
\begin{equation}\label{eq1}
ds^2=-dt^2+a^2(t)(\frac{1}{1-Kr^2}dr^2+r^2d\Omega^2),
\end{equation}
where $K=+1, 0, -1$ denotes the spatial curvature for closed,
flat and open geometries, respectively. Note that the curvature parameter is directly related to the
constant $K$ as $\Omega_k=-k/a_0^2 H_0^2$. Now we respectively denote the dimensionless comoving distances
$d_l\equiv d(0, z_l)$, $d_s\equiv d(0, z_s)$ and $d_{ls}\equiv d(z_l, z_s)$, in the framework of
strong lensing system with a source (at redshift $z_s$) observed on the image plane (at redshift $z_l$).
Note that the dimensionless comoving distances ($d$)
\begin{equation}
d(z_l, z_s)=\frac{1}{\sqrt{|\Omega_k|}}S_K\left(
\sqrt{|\Omega_k|}\int^{z_s}_{z_l}\frac{H_0dz'}{H(z')} \right),
\end{equation}
where
\begin{equation}\label{eq3}
S_K(x)=\left\{
   \begin{array}{lll}
   \sin(x)\qquad \,\ \Omega_k&<0, \\
   x\qquad\qquad \,\ \Omega_k&=0, \\
   \sinh(x)\qquad \Omega_k&>0. \\
   \end{array}
   \right.
\end{equation}
are connected with the angular diameter distance ($D_A$) as $d(z_l, z_s)= (1+z_s)H_0 D_A(z_l,z_s)$. As was originally proposed in \citet{Ratra88}, the distance sum rule in non-flat FLRW models gives
\begin{equation}\label{eq4}
d_{ls}={d_s}\sqrt{1+\Omega_kd_l^2}-{d_l}\sqrt{1+\Omega_kd_s^2}.
\end{equation}
with $d'(z)>0$ and a one-to-one correspondence between the cosmic
time $t$ and redshift $z$ \citet{Bernstein06}. Such simple relation,
which was first used to obtain model-independent measurements of the
spatial curvature \citep{Clarkson2008,Qi2019}, has also been
recently discussed to test the validity of the FLRW metric in the
Universe \citep{Qi2019b} based on different types of gravitational
lensing events. Now we could rewrite this fundamental relation so
that the strong lensing observations (from LSST lenses) and
luminosity distances (from DECIGO standard sirens) are encoded
\begin{equation} \label{smr}
\frac{d_{ls}}{d_s}=\sqrt{1+\Omega_Kd_l^2}-\frac{d_l}{d_s}\sqrt{1+\Omega_K
d_s^2}.
\end{equation}
The left item can be derived from the source/lens distance ratio
$d_{ls}/d_s=D^{A}_{ls}/D^{A}_s$, based on high-resolution imaging
and spectroscopic observations in SGL systems \citep{Cao2015b}.

For a strong lensing system with early-type galaxy acting as
intervening lens, one of its typical feature is the Einstein radius
($\theta_E$) depends on the source/lens distance ratio
($d_{ls}/d_s$), the lens velocity dispersion ($\sigma$), and the
density profiles of the lens galaxies. Such methodology was
originally proposed in \citet{Futamase01} and extended to different
SGL samples \citep{Bolton08,Cao2012,Cao2015b,Chen19}, with the aim
of quantitatively studying the redshift evolution of cosmic equation
of state \citep{Li16,Liu19}, measuring the speed of light at
different redshifts \citep{Cao18,Cao20}, and testing the General
Relativity at large scale \citep{Cao17c,Collett18}. In this paper,
three different models will be included in our analysis to describe
the mass distribution of early-type galaxies: Singular Isothermal
Ellipsoid (SIE) lens model, Power-law lens model, and Extended
power-law model \citep{Chen19,Zhou20}. If the lens mass profile can
be approximately described by SIE, the distance ratio is expressed
as \citep{Koopmans06}
\begin{equation} \label{SIE_E}
\frac{d_{ls}}{d_s}=\frac{c^2\theta_E}{4\pi \sigma_{SIE}^2}
=\frac{c^2\theta_E}{4\pi \sigma_{0}^2f_E^2},
\end{equation}
where $\sigma_{SIS}$ and $c$ respectively denote the SIS (Singular
Isothermal Sphere) velocity dispersion and the speed of light. The
parameter $f_E$ is introduced to quantify different systematics that
could change the observed multiple image separation or generate the
difference between $\sigma_{SIS}$ and the observed velocity
dispersion of stars ($\sigma_{0}$). The relation between the
measurement of $\sigma_{0}$ from spectroscopy and that estimated
from the SIE model has been extensively discussed in
\citet{Ofek2003,Cao2012}. In the second case, we take into account a
spherically symmetric power-law mass distribution ($\rho\sim
r^{-\gamma}$,  $r$ is the spherical radial coordinate from the lens
center,) to generalize the simplest Singular Isothermal Sphere lens
model, considering the non-negligible deviation from SIS
($\gamma=2$) based on recent observations of  the density profiles
of early-type galaxies \citep{Koopmans06,Humphrey10,Sonnenfeld13a}.
Now the corresponding distance ratio is rewritten as
\citep{Ruff2011,Koopmans06,Bolton2012}
\begin{equation} \label{sigma_gamma}
\frac{d_{ls}}{d_s}=\frac{c^2\theta_E}{4\pi
\sigma_{ap}^2}\left(\frac{\theta_{ap}}{\theta_E}\right)^{2-\gamma}f^{-1}(\gamma),
\end{equation}
where $f(\gamma)$ is a function of the radial mass profile slope
($\gamma$) and $\sigma_{ap}$ denotes the projected, luminosity
weighted average of the velocity dispersion inside the circular
aperture $\theta_{ap}$ (See \citet{Cao2015b} for the derivation of
equivalent $\sigma_{ap}$ within rectangular apertures). One of the
advantages of such power-law lens model is based on the assumption
that the distribution of stellar mass follows the same power law as
that of the total mass, with the vanishing of velocity anisotropy
\citep{Koopmans05}. Consequently, we take into account these
uncertainties by introducing a general mass model for the early-type
lens galaxies, with the total (i.e. luminous plus dark-matter) mass
density distribution ($\rho(r)\sim r^{-\alpha}$) and the luminosity
density profile ($\nu(r)\sim r^{-\delta}$). Here we choose to
consider the anisotropy of the stellar velocity dispersion in this
analysis, which is quantified by the a new parameter $\beta(r) = 1 -
{\sigma^2_\theta} / {\sigma^2_r}$, where $\sigma_\theta$ and
$\sigma_r$ are the tangential and radial velocity dispersions,
respectively. In the framework of such extended power-law model, the
distance ratio can be computed from the radial Jeans equation in
spherical coordinate system \citep{Koopmans06}, by projecting the
dynamical mass to the lens mass within the Einstein radius
\begin{eqnarray}\label{sigma_alpha_delta}
\nonumber
\frac{d_{\rm ls}}{d_{\rm s}}&=& \left(\frac{c^2}{4\sigma_{ap}^2}\theta_{\rm E}\right)\frac{2(3-\delta)}{\sqrt{\pi}(\xi-2 \beta)(3-\xi)} \left( \frac{\theta_{\rm ap}}{\theta_{\rm E}}\right)^{2-\alpha}\\
&\times&\left[\frac{\lambda(\xi)-\beta\lambda(\xi+2)}{\lambda(\alpha)\lambda(\delta)}\right]~,
\end{eqnarray}
where $\xi=\alpha+\delta-2$,
$\lambda(x)=\Gamma(\frac{x-1}{2})/\Gamma(\frac{x}{2})$. It is
apparent that this extended power-law model will reduce to the
power-law lens model when $\delta=\alpha$, i.e., the distribution of
stellar mass follows the same power law as that of the total mass.

Combing the above equations with the error propagation formula Eq.~(6), we could obtain the uncertainty of SGL systems
($\sigma_{SGL}$) for different lens models, based on the
observational uncertainties of the Einstein radius and velocity
dispersion. The distance information $d(z)$ in the right items of
Eq.~(\ref{smr}) is determined by luminosity distances from
gravitational wave data. In this paper, we will present an updated
estimation of the cosmic curvature or the FLRW metric from the
largest SGL sample by LSST and future GW observations by DECIGO.

\begin{figure*}
\begin{center}
\includegraphics[width=0.45\linewidth]{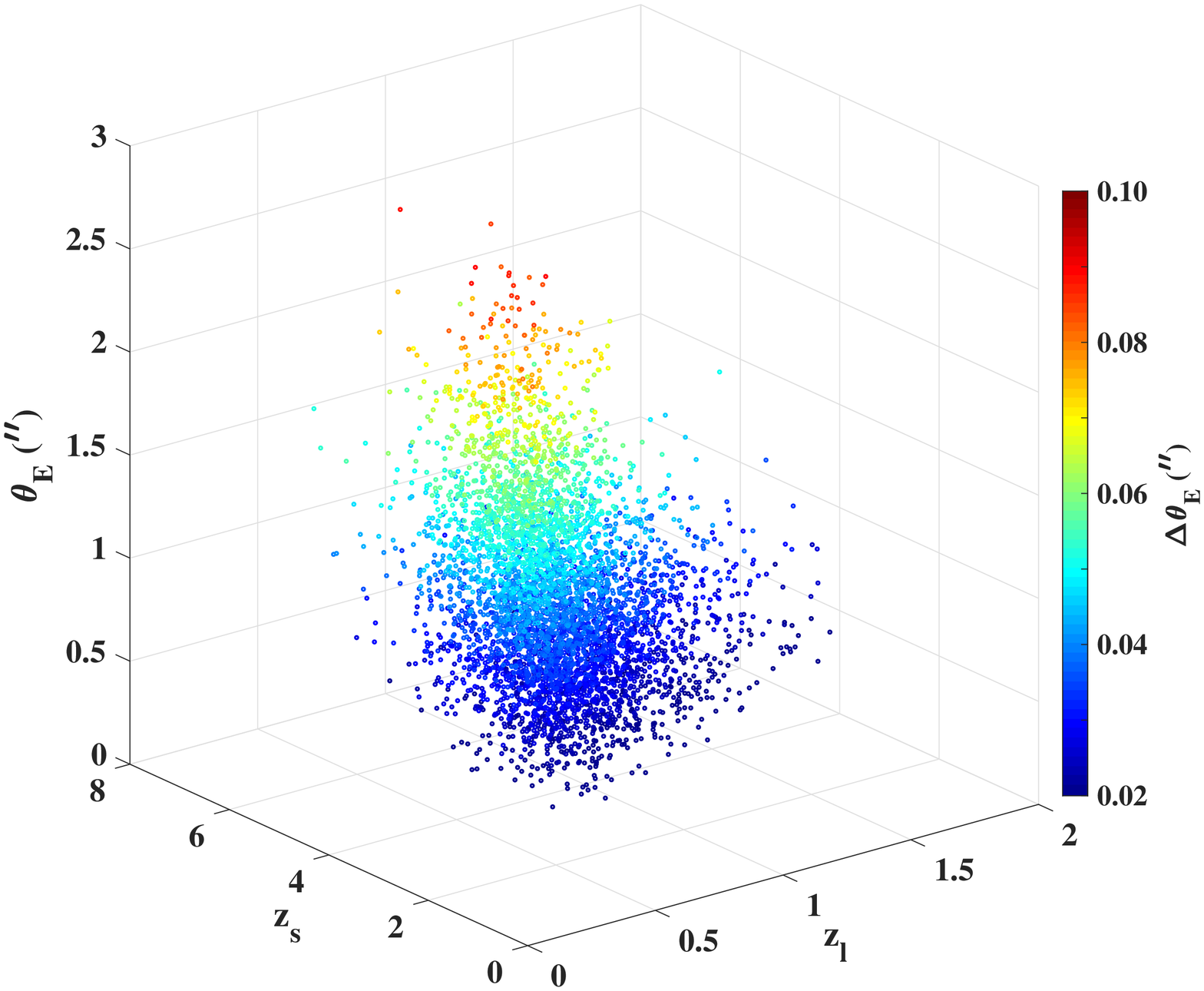}
\includegraphics[width=0.45\linewidth]{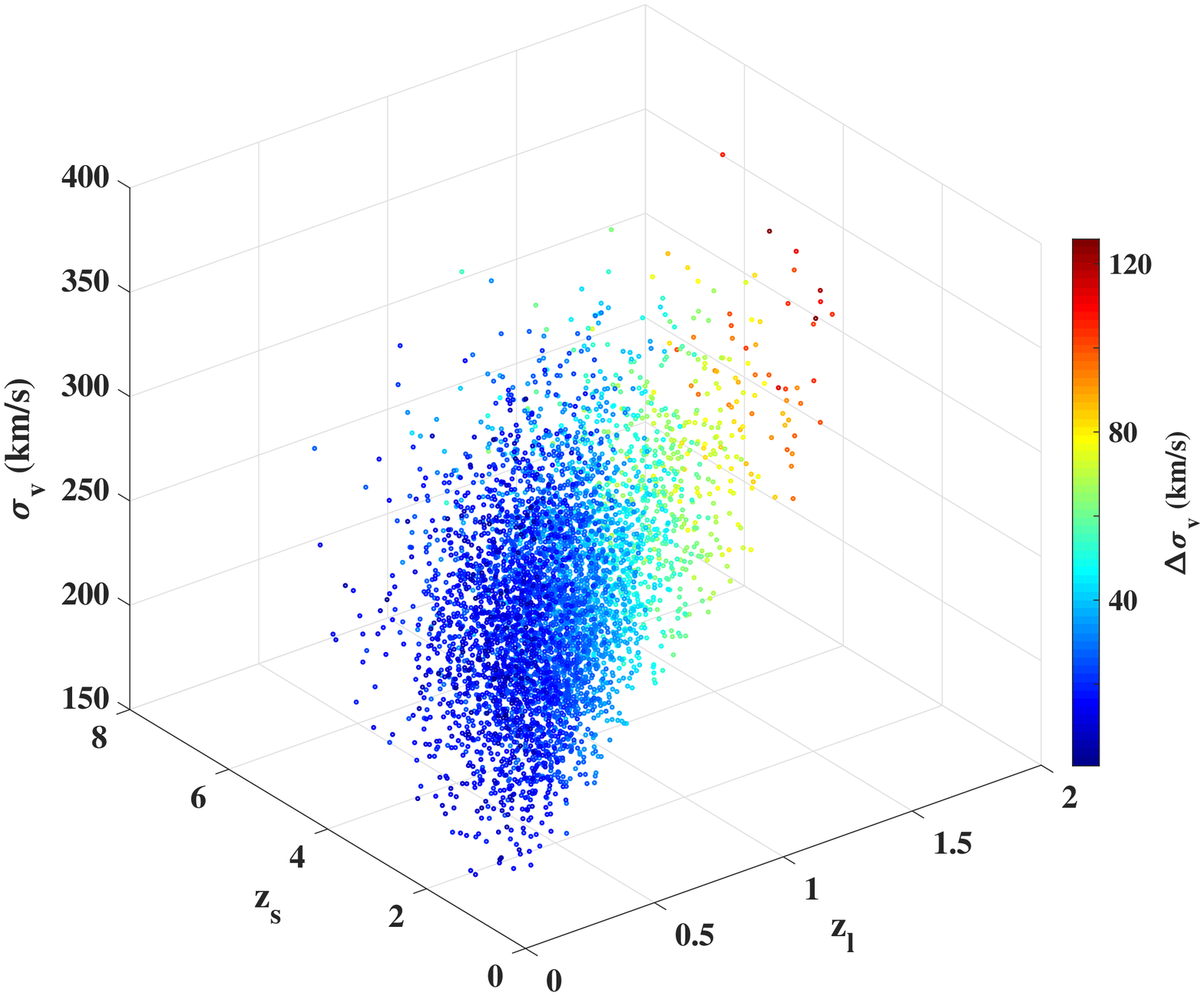}
\end{center}
\caption{The scatter plot of the simulated LSST lensing systems,
with the gradient color denoting the uncertainty the Einstein radius
and lens velocity dispersion. }
\end{figure*}

\section{Simulations from DECIGO and LSST}

\subsection{Strong lenses from LSST}

As one of the most important wide-area and deep surveys besides the
Dark Energy Survey (DES) \citep{Frieman2004}, the upcoming Large
Synoptic Survey Telescope (LSST) is expected to monitor $\sim 10^5$
strong gravitational lenses in the most optimistic discovery
scenario, by repeatedly scanning nearly half of the sky for ten
years \citep{Collett15}. Such tremendous increase of known
galaxy-scale lenses by orders of magnitude will produce extensive
cosmological applications in the near future
\citep{Cao17c,Cao18,Ma19,Cao20}. With high-quality imaging and
spectroscopic data, the Einstein radius of multiple images and the
lens velocity dispersion can be measured precisely and accurately.

In order to assess the performance of forthcoming optical imaging
surveys, the simulation of a realistic population of galaxy-galaxy
strong lenses has been performed \citep{Collett15}. The results
showed that although $\sim 10^5$ strong gravitational lenses are
discoverable in LSST, only a fraction of SGL sub-sample is available
for our curvature estimation, given the expensiveness of substantial
follow-up efforts and dedicated spectroscopic observations
(spectroscopic velocity dispersion, spectroscopic confirmation of
the lens and source redshift) \citep{Hlozek19}. Therefore, in this
paper we will simulate a particularly well-selected sub-sample of
LSST lenses with the observations of the foreground deflector and
the background source population
\footnote{github.com/tcollett/LensPop}, following the recent
analysis of multi-object and single-object spectroscopy to enhance
Dark Energy Science from LSST \citep{Mandelbaum19}. More
specifically, it is more realistic to focus only on 5000
well-measured systems with intermediate-mass early-type galaxies
acing as strong gravitational lenses, with the velocity dispersion
of 200 km/s $<\sigma_{ap} <$ 300 km/s. Such criteria is strongly
supported by the recent findings that for systems with velocity
dispersion between 200 km/s to 300 km/s, there is a good consistency
between the measurement of $\sigma_0$ from spectroscopy and those
estimated from the SIS lens model \citep{Treu06,Cao2016}. In our
simulations, we choose the velocity dispersion function (VDF) from
DSS Data Release 5 \citep{Choi07} to describe the number density of
these lensing galaxies, the mass distributions of which are well
quantified by the singular isothermal sphere (SIS) model. We take
the fractional uncertainty of the Einstein radius and the observed
velocity dispersion following the uncertainty budget proposed in
\citet{Liu20}.

To assess the analysis of Einstein radius extraction from future
LSST survey, some recent attempts have been made to investigate the
effect of line-of-sight contamination \citep{Hilbert09,Collett16},
which found that for monitorable strong lenses such effect could
introduce a 1-3\% uncertainty in the Einstein radius measurements.
Such error strategy has been extensively used in the simulation of
LSST lens sample, with high-quality (sub-arcsecond) imaging data in
general \citep{Cao17c}. Based on the observations of 32 strong
lensing systems from Strong Lensing Legacy Survey (SL2S), with both
Canada-France-Hawaii Telescope (CFHT) near-infrared ground-based
images or Hubble Space Telescope (HST) imaging data
\citep{Sonnenfeld13}, \citet{Liu20} recently investigated the
possible anti-correlation between the fractional uncertainty of the
Einstein radius ($\Delta \theta_E$) and $\theta_E$. The results
showed that different error strategies should be applied to strong
lenses with different Einstein radii ($\theta_E$), due to the fact
that strong lensing systems with smaller $\theta_E$ will be
accompanied with larger statistical uncertainties. Therefore,
different from the previous work which took a constant precision for
each SGL system observed with HST-like image quality \citep{Cao17c},
we take 8\%, 5\% and 3\% as the average Einstein radius precision
for each system, which could be classified as small Einstein radii
lenses ($0.5"<\theta_E<1.0"$), intermediate Einstein radii lenses
($1"\leq\theta_E<1.5"$), and large Einstein radii lenses
($\theta_E\geq1.5$) with HST+CFHT imaging data.

Moreover, \citet{Liu20} recently proposed that other intrinsic
properties of the lensing system (such as the total mass or the
brightness of the lensing galaxy) could significantly change the
observational precision of lens velocity dispersion. The lessons in
the statistical analysis of 70 intermediate-mass lenses (with
average velocity dispersion of $\sigma_{ap}\sim 230$ km/s) observed
by Sloan Lens ACS survey (SLACS) \citep{Bolton08} showed the
fractional uncertainty is strongly correlated with the lens surface
brightness in the $i$-band. To incorporate this effect, we consider
the anti-correlation between these two quantities and take the
best-fitted correlation function obtained in \citet{Liu20} to
simulate the velocity dispersion uncertainty for each LSST lens.
Note that such strategy is different from that of the previous work,
which assigned an overall error of 5\% on each SGL system observed
with detailed follow-up spectroscopic information from other
ground-based facilities \citep{Cao2015b,Zhou20}. The Einstein radius
and velocity dispersion distributions of the simulated LSST lenses
are plotted in Fig.~1.

\begin{figure}
\begin{center}
\includegraphics[width=0.95\linewidth]{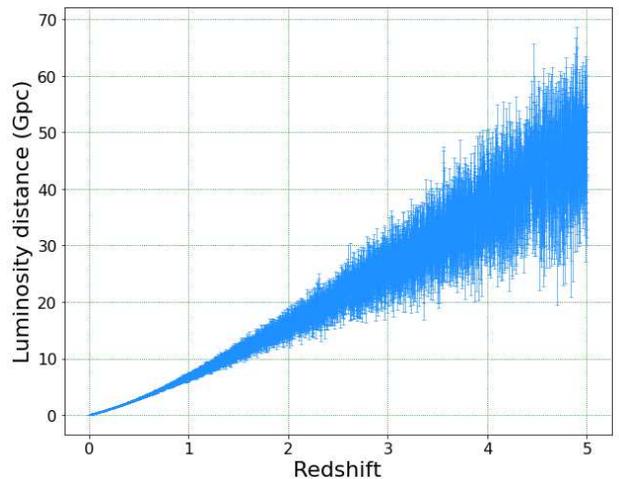}
\end{center}
\caption{The luminosity distance measurements from 10,000 simulated
GW events observable by the space detector DECIGO.}
\end{figure}

\begin{figure}
\begin{center}
\includegraphics[width=0.95\linewidth]{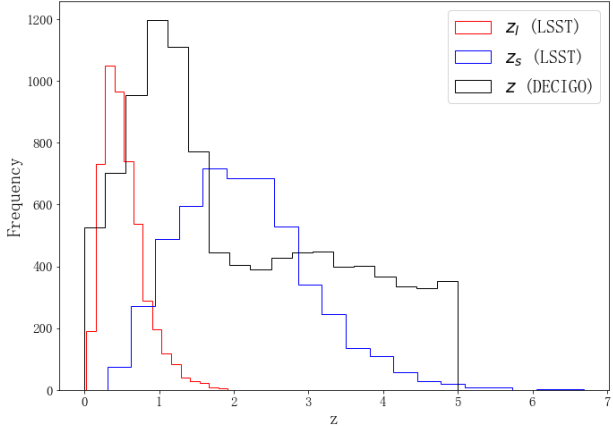}
\end{center}
\caption{Redshift distributions for the GWs from DECIGO, and strong
lensing systems (including the source and the lens) from LSST.}
\end{figure}

\begin{table*}
\begin{center}
\begin{tabular}{c| c c c c c c c}
\hline
   & $\Omega_k$ & $f_E$ &$\gamma_0$&$\gamma_1$ &$\alpha$ & $\delta$ &$\beta$  \\
\hline
 SIS   & $0.0001^{+0.012}_{-0.013}$ & $1.000^{+0.002}_{-0.002}$ & $\Box$ & $\Box$ & $\Box$ & $\Box$ & $\Box$\\
\hline
 Power-law spherical  & $-0.016^{+0.035}_{-0.037}$ & $\Box$  &$2.001^{+0.001}_{-0.001}$ & $0.002^{+0.003}_{-0.003}$ & $\Box$  & $\Box$ & $\Box$ \\
\hline
Extended power-law   & $-0.007^{+0.050}_{-0.047}$ & $\Box$ & $\Box$  & $\Box$ &$2.003^{+0.011}_{-0.011}$ & $1.968^{+0.527}_{-0.516}$ &$-0.067^{+0.605}_{-0.325}$ \\
\hline

\end{tabular}
\caption{Summary of constraints on the cosmic curvature $\Omega_k$
and lens model parameters for three types of lens models (See the
text for definitions)} \label{SIE_table}
\end{center}
\end{table*}

\subsection{Standard sirens from DECIGO}

It is well known that GW from inspiraling binaries could provide a
new, independent probe of the cosmic expansion using standard sirens
\citep{Abbott16,Abbott17}. More importantly, the standard siren
method, which focuses on binary neutron star mergers coupled with
electromagnetic (EM) measurements of the redshift, has already been
used to produce constraints on the Hubble constant at lower redshift
\citep{Zhang20}, the cosmic opacity and distance duality relation at
much higher redshifts \citep{Qi2019b,Qi2019c}.

Using laser interferometry, the DECi-Hertz Interferometric
Gravitational Observatory (DECIGO) is a space mission designed to
open the DECi-Hertz frequency range to GW observations
\citep{Seto01,Kawamura11}. Compared with the ground-based detectors
and other space-based detector such as Laser Interferometric Space
Antenna (LISA), DECIGO is expected to detect different populations
of GW sources in this unique frequency range of 0.1-1 Hz, including
primordial black holes, intermediate-mass black hole binary
coalescences, neutron star binary coalescences, and black hole-
neutron star binaries in their inspiral phase \citep{Kawamura11}.
The loudest objects in this band of the GW sky are expected to be
mergers of neutron star binary coalescences, long time before they
enter the LIGO frequency range. Such advantage, which significantly
increases the precision of inferences made from chirp signals
\citep{Ola20}, which yield orders of magnitude more candidate
standard sirens in the earlier period of the Universe. More
specifically, following the recent estimation of \citet{Kawamura19},
DECIGO will bring us the yearly detection of 10,000 NS-NS systems at
redshift $z\sim$5, based on the frequency of the binary coalescences
given above. Note that although GW may provide us some information
about the source redshifts \citep{Messenger12,Messenger14},
observations of the EM counterparts or host galaxies by ground-based
telescopes are still necessary for these expected GW signals
\citep{Cutler09}. This offers the exciting possibility we explore:
the ability of deep-redshift standard sirens observed by DECIGO to
validate or refute the flat geometry inferred by the newest Planck
observations.

Following the simulation process by \citet{Geng20}, we generate mock
DECIGO NS-NS standard siren observations based on a flat
$\Lambda$CDM cosmology and assume that their redshift is known. The
NS mass distributions is chosen uniformly in [1,2] $M_\odot$. For
each coalescing NS-NS system with physical masses ($m_1$ and $m_2$)
and symmetric mass ratio ($\eta=m_{1}m_{2}/M_{t}^{2}$), one could
derive the Fourier transform of the GW waveform as
\begin{equation}
    \widetilde{h}(f)=\frac{A}{D_{L}(z)}M_{z}^{5/6}f^{-7/6}e^{i\Psi(f)},
\end{equation}
where $A=(\sqrt{6}\pi^{2/3})^{-1}$ quantifies the geometrical
average over NS-NS system's inclination angle, while $D_{L}(z)$
denotes the luminosity distance to the source with the redshifted
chirp mass of $M_{z}=(1+z)\eta^{3/5}M_{t}$. As proposed in previous
work \citep{Maggiore08}, the frequency-dependent phase caused by
orbital evolution ($\Psi(f)$) can be derived from 1.5 (or higher)
post-Newtonian (PN) approximation. For the purpose of uncertainty
estimation, different sources of uncertainties are included in our
simulation of luminosity distance. On the one hand, focusing only on
the inspiral phase of the GW signal, the instrumental uncertainty
for a nearly face-on case is given by
$\sigma^{inst}_{D_{L,GW}}=\frac{2D_{L,GW}}{\rho}$, where $\rho$ is
the combined SNR for the network of space-borne detector and the
factor of 2 is included to quantify the maximal effect of the
inclination on the SNR \citep{Zhao11}. On the other hand, the
lensing uncertainty caused by the weak lensing is the other crucial
point of our idea, due to large scale structure that could
potentially bias the results especially at high redshifts
\citep{Sathyaprakash2010}. More specifically, following the
procedure extensively applied in the literature \citep{Zhang20}, it
was recently proposed that such weak lensing uncertainty could be
modeled as $\sigma^{lens}_{D_{L}}/D_{L}=0.044z$ for the space-based
GW detectors \citep{Cutler09}. Therefore, the total uncertainty on
the luminosity distance is given by
\begin{eqnarray}
\sigma_{D_{L,GW}}&=&\sqrt{(\sigma_{D_{L,GW}}^{\rm inst})^2+(\sigma_{D_{L,GW}}^{\rm lens})^2} \nonumber\\
            &=&\sqrt{\left(\frac{2D_{L,GW}}{\rho}\right)^2+(0.05z D_{L,GW})^2}.
\label{sigmadl}
\end{eqnarray}
With the luminosity distance from the standard sirens, the
uncertainty in $\sigma_{GW}$ can be expressed as a function of
$D_{L,GW}^s$, $D_{L,GW}^l$, $\sigma_{D_{L,GW}}^s$, and
$\sigma_{D_{L,GW}}^l$ through the error propagation formula
[Eq.~(5)].

Now the final key question required to be answered is: how to
describe the redshift distribution of GW events that can be detected
by DECIGO? Given the analytical fit of DECIGO noise spectrum
including the shot noise, the radiation pressure noise and the
acceleration noise \citep{Kawamura19,Kawamura06,Nishizawa10,Yagi11},
the simulated luminosity distances from 10,000 standard sirens in
DECIGO is presented in Fig.~2, with the redshift distribution
follows the form provided by
\citet{Sathyaprakash2010,Cutler06,Schneider01}. We refer to
\citet{Geng20} for more simulation details of DECIGO standard
sirens. For a good comparison, Fig.~3 illustrates the perfect
redshift coverage of the simulated DECIGO sample, compared with
lensing observations including the source and lens from LSST.

\section{Results and discussion}

In this section, we describe the observational constraint on the
cosmic curvature using the observational data-set summarized in
Sect. 3. In particular, we simultaneously fit the spatial curvature
parameter and lens model parameters to the LSST lens sample and
luminosity distance data from DECIGO, and find the best-fit of
$\Omega_K$ in the DSR. The statistical quantity $\chi^2$ is written
as
\begin{equation}
\chi^2(\textbf{p},\Omega_k)=\sum_{i=1}^{N} \frac{\left({\cal
D}_{GW}({z}_i;\Omega_k)- {\cal
D}_{SGL}({z}_i;\textbf{p})\right)^2}{\sigma_{\cal D}(z_i)^2},
\end{equation}
with the two factors contributing to the uncertainty of distance
ratio from the observables of the strong lensing systems in LSST and
the luminosity distance measurements from standard sirens in DECIGO.
We assume that the two uncertainties of LSST lenses and DECIGO
standard sirens are uncorrelated and thus they add in quadrature of
$\sigma_D^2=\sigma_{SGL}^{2}+\sigma_{QSO}^2$. In order to calculate
the posterior distribution of the model parameters, we use the
Python module emcee
\footnote{https://emcee.readthedocs.io/en/stable/}, which is an
Affine Invariant Markov chain Monte Carlo (MCMC) Ensemble sampler
\citep{Foreman13}, to survey the posterior distribution in parameter
space and to maximize the likelihood function ${\cal L} \sim
\exp{(-\chi^2 / 2)}$.

We assume three kinds of spherically symmetric mass distributions
(SIS, power-law model, and extended power-law model) for the lensing
galaxies in the cosmic curvature analysis. The 1D and 2D
marginalized distributions with 1$\sigma$ and 2$\sigma$ confidence
level contours for $\Omega_K$ and relevant lens parameters
constrained from the combined LSST+DECIGO data are shown in
Fig.~4-6.

\begin{figure}
\begin{center}
\includegraphics[width=0.95\linewidth]{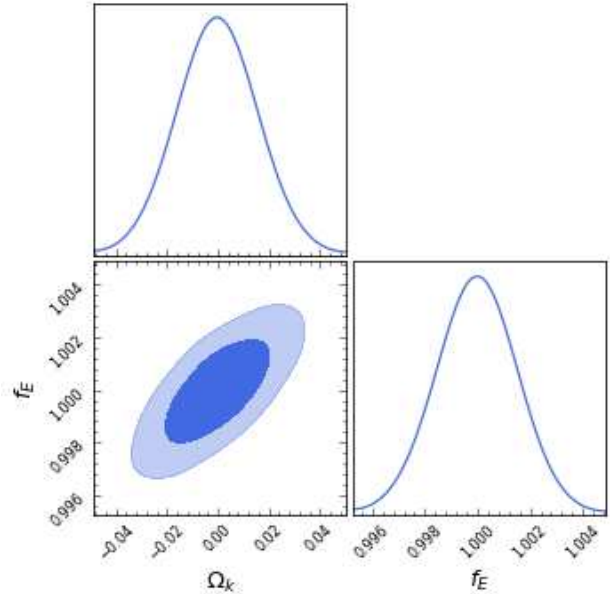}
\end{center}
\caption{The 2-D regions and 1-D marginalized distribution with the
1-$\sigma$ and 2-$\sigma$ contours of all parameters, in the
framework of SIS lens models.}
\end{figure}

\begin{figure}
\begin{center}
\includegraphics[width=0.95\linewidth]{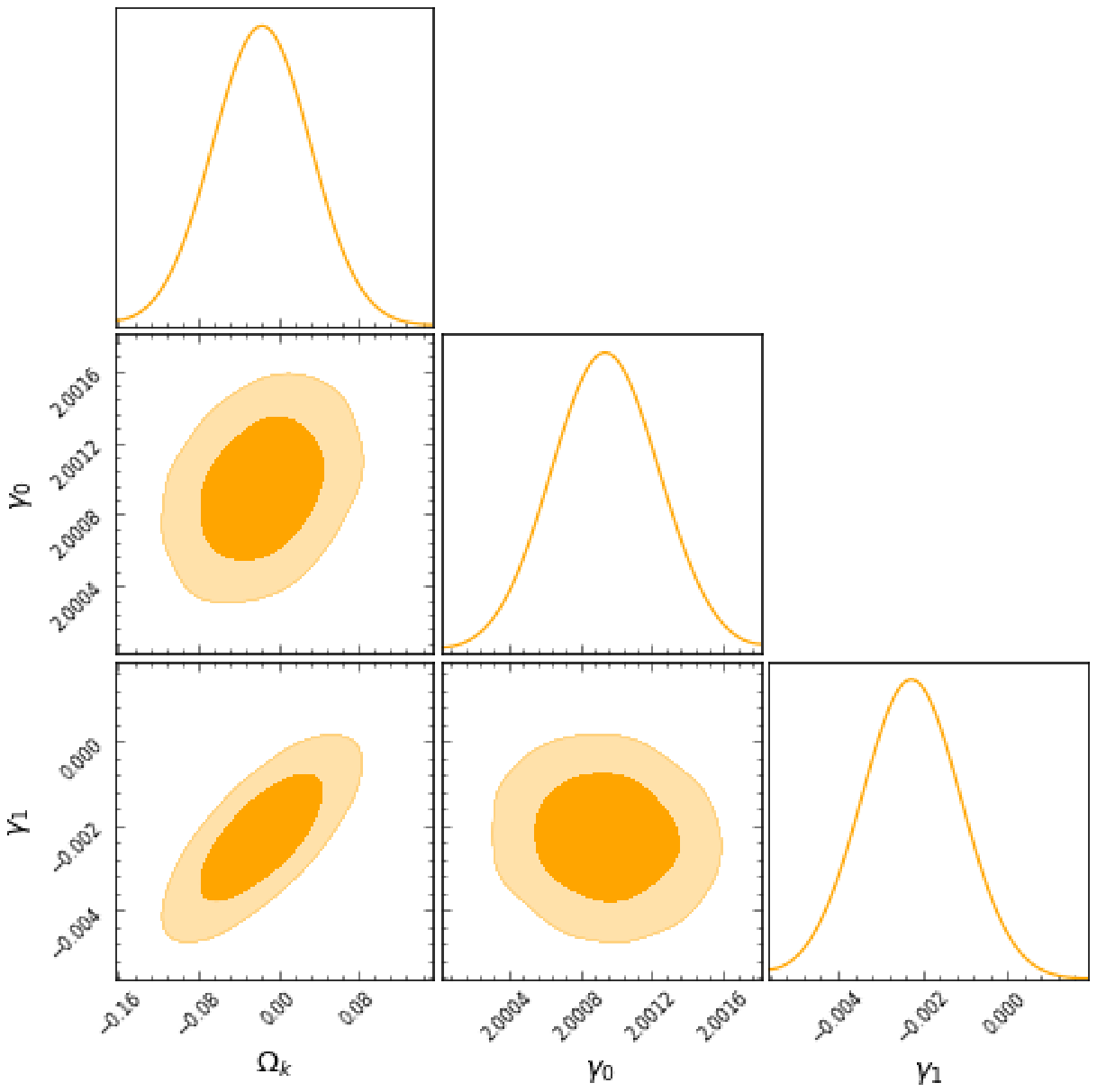}
\end{center}
\caption{The 2-D regions and 1-D marginalized distribution with the
1-$\sigma$ and 2-$\sigma$ contours of all parameters, in the
framework of power-law lens profile.}
\end{figure}

\begin{figure}
\begin{center}
\includegraphics[width=0.95\linewidth]{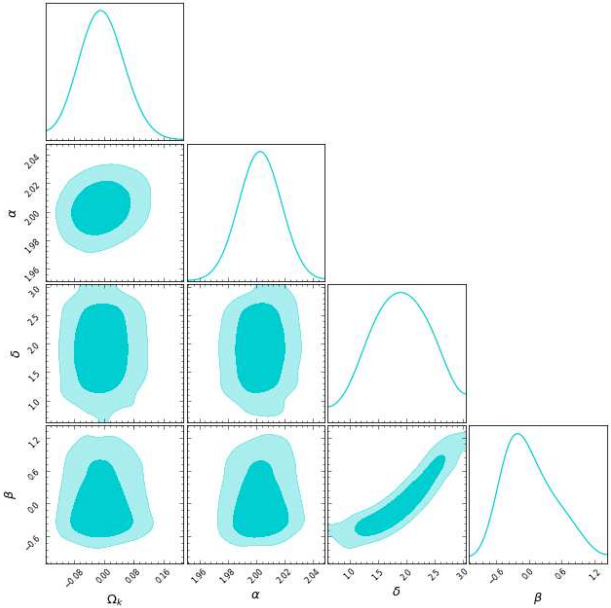}
\end{center}
\caption{The 2-D regions and 1-D marginalized distribution with the
1-$\sigma$ and 2-$\sigma$ contours of all parameters, in the
framework of extended power-law lens profile.}
\end{figure}

In the framework of the simplest SIS model for the first application
of the method described above, we obtain the best-fit value of
curvature parameter and the corresponding $1\sigma$ (more precisely
68\% confidence level) uncertainties:
$\Omega_K=0.0001^{+0.012}_{-0.013}$, which indicates that the cosmic
curvature can be accommodated at very high precision ($\Delta
\Omega_K=10^{-2}$) comparable to that derived from the Planck CMB
power spectra \citep{Planck Collaboration}. Compared with the
previous results obtained in \citet{Rasanen2015}, the fits on the
cosmic curvature have been improved by two orders of magnitudes,
benefit from the significant increase of well-measured strong
lensing systems and standard sirens in the future. Such conclusion
could also be derived through the comparison with other works by
using different available SGL sub-samples \citet{Xia2017,Qi2019}. In
this analysis, the lens parameter characterizing the mass
distribution profile $f_E$ is also estimated in a global fitting
without taking any prior. To study the correlation between
$\Omega_K$ and $f_E$, we display the two-dimensional (2D)
probability distributions in the ($\Omega_K$, $f_E$) plane, with the
marginalized $1\sigma$ constraint of the parameter
$f_E=1.000^{+0.002}_{-0.002}$. It is interesting to note that our
results, which strongly support a flat universe together with the
validity of SIS model ($\Omega_K=0$, $f_E=1$), reveal significant
degeneracy between the spatial curvature of the universe and the
$f_E$ parameter, which characterizes the mass distribution of the
lensing galaxies. The numerical results are also summarized in Table
1.

Now we focus our attention on the constraints on the parameters in
the framework of power-law mass density profile, in which the mass
density power-law index of massive elliptical galaxies evolves with
redshift ($\gamma=\gamma_0+\gamma_1\times z_l$). Performing fits on
the data comprising the strong lensing systems in LSST and the
luminosity distance measurements from standard sirens in DECIGO, we
obtain the following best-fit values and corresponding $1\sigma$
uncertainties: $\Omega_K=-0.016^{+0.035}_{-0.037}$,
$\gamma_0=2.001^{+0.001}_{-0.001}$ and
$\gamma_1=0.002^{+0.003}_{-0.003}$. The 1D and 2D marginalized
distribution for $\Omega_K$ and the power-lens model parameters are
shown in Fig.~5. It is turned out that the fit on the cosmic
curvature in this case is still very strong compared with that found
for the SIS model, which indicates that our findings are quite
robust. Meanwhile, contour plots in Fig. 5 show that the
degeneracies among the cosmic curvature and lens parameters
($\gamma_0$) in power-law mass density profile are similar to that
of the SIS model ($f_E$). More specifically, it is easy to find from
the $\Omega_K-\gamma_1$ contour that $\Omega_K$ is strongly
correlated with $\gamma_1$, which indicates that a significant
redshift evolution of the mass density power-law index will result
in a larger cosmic curvature. Therefore, compared with the previous
results focusing on a constant power-law lens index parameter
\citep{Zhou20}, our analysis reveals that the estimation of the
spatial curvature is more sensitive to the measurement of its
possible evolution with redshifts. Such conclusion, which has been
suggested by \citet{Ruff2011}, is also further supported by
\citet{Sonnenfeld13a} in a combined sample of lenses from SLACS,
SL2S, and LSD. Therefore, additional observational information, such
as dedicated follow-up imaging of the lensed images for a sample of
individual lenses is necessary in this case. Such high-cadence,
high-resolution and multi-filter imaging could be technically
obtained through frequent visits on Hubble telescope or smaller
telescopes on the ground \citep{Collett14,Wong15}.

Let us finally focus on the performance of the extended power-law
lens model, in which the total density slope, luminosity density
slope and the anisotropy of stellar velocity dispersion are taken as
free parameters. The graphical and numerical results from the
combined DECIGO+LSST data set are displayed in Fig.~6 and Table 1.
The values listed in Table 1 show that the extended power-law lens
model generates competitive constraints on the cosmic curvature
($\Omega_K=-0.007^{+0.050}_{-0.047}$) comparable with the power-law
mass density profile. This is inconsistent with the results
presented in \citet{Xia2017}, where they have used the similar lens
model (without considering the effect of the anisotropy of stellar
velocity dispersion) but but adopted different combinations of
available data in the EM domain. Compared with the SNe Ia standard
candles, the advantage of DECIGO standard sirens is that larger
number of GWs could be observed at much higher redshifts, which
motivate us to calibrate the LSST strong lenses and investigate the
cosmic curvature in the early universe. More interestingly, the
extended power-law lens model is more suitable to estimate the mass
distribution of baryonic matter and dark matter in the early-type
galaxies. Compared with the previous observational constraints on
the total-mass density profile \citep{Cao2016,Xia2017,Chen19}, the
combined LSST+DECIGO data will improve the constraints significantly
($\alpha=2.003^{+0.011}_{-0.011}$ with respect to current results,
showing the high constraining power that can be reached by the
forthcoming surveys. Furthermore, as can be seen in Fig.~6 the
addition of DECIGO to the combination of LSST+DECIGO does improve
the constraint on the luminosity density slope
($\delta=1.968^{+0.527}_{-0.516}$) and the anisotropy of the stellar
velocity dispersion significantly
($\beta=-0.067^{+0.605}_{-0.325}$). Such steeper luminosity density
profile and nonzero stellar velocity anisotropy parameter are
consistent with \citet{Cao2016,Zhou20} and also with recent results
of Illustris simulations \citep{Xu16}, focusing on early-type
galaxies with spherically symmetric density distributions. In this
case, auxiliary data such as integral field unit (IFU) spectroscopic
data, especially Adaptive optics (AO) IFU spectroscopy on
8-40m-class telescopes or AO imaging with slit spectroscopy
\citep{Hlozek19}, could provide complementary information of
$\delta$ and $\beta$ in the near future \citep{Barnabe13}.

\begin{figure*}
\begin{center}
\includegraphics[width=0.4\linewidth]{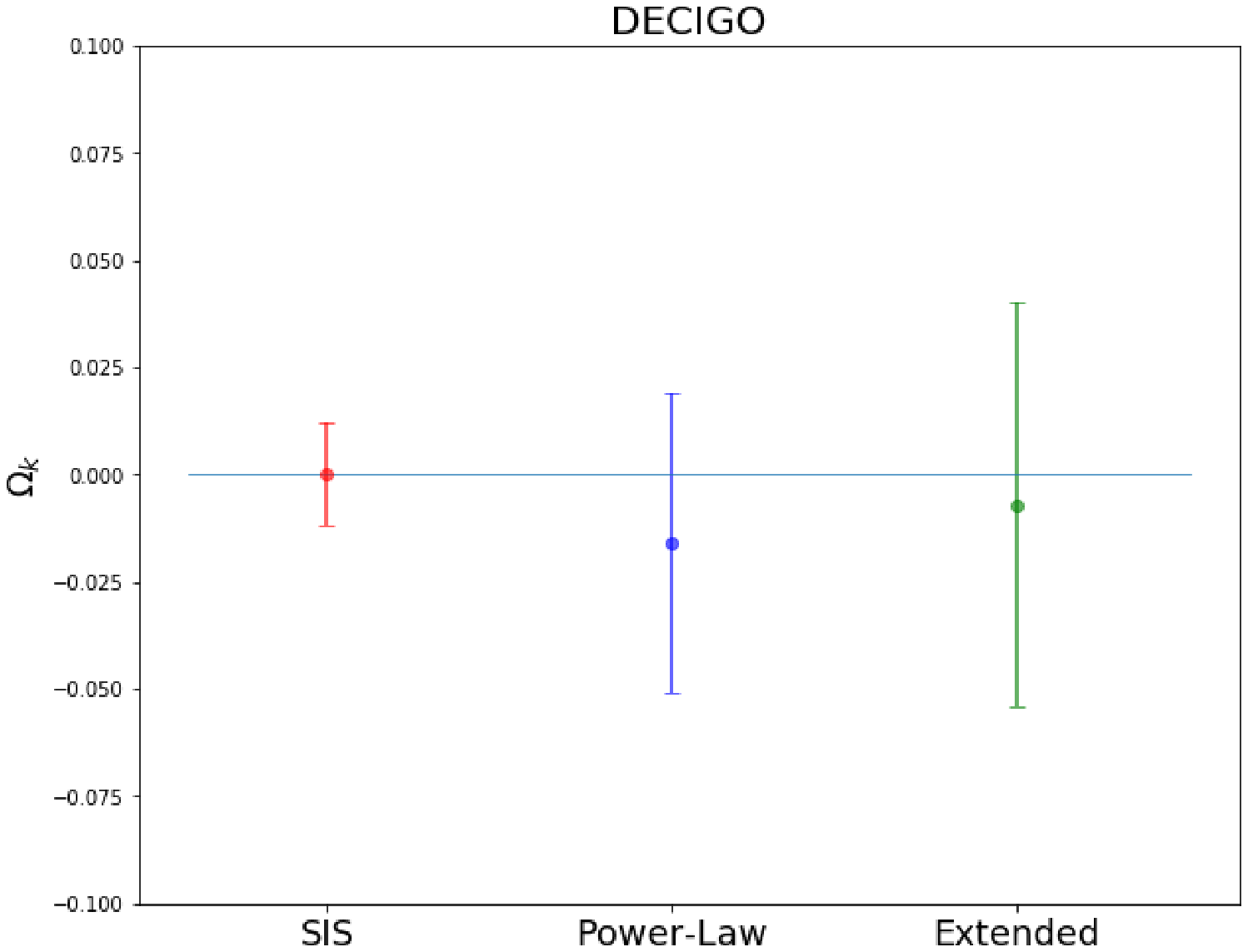} \includegraphics[width=0.4\linewidth]{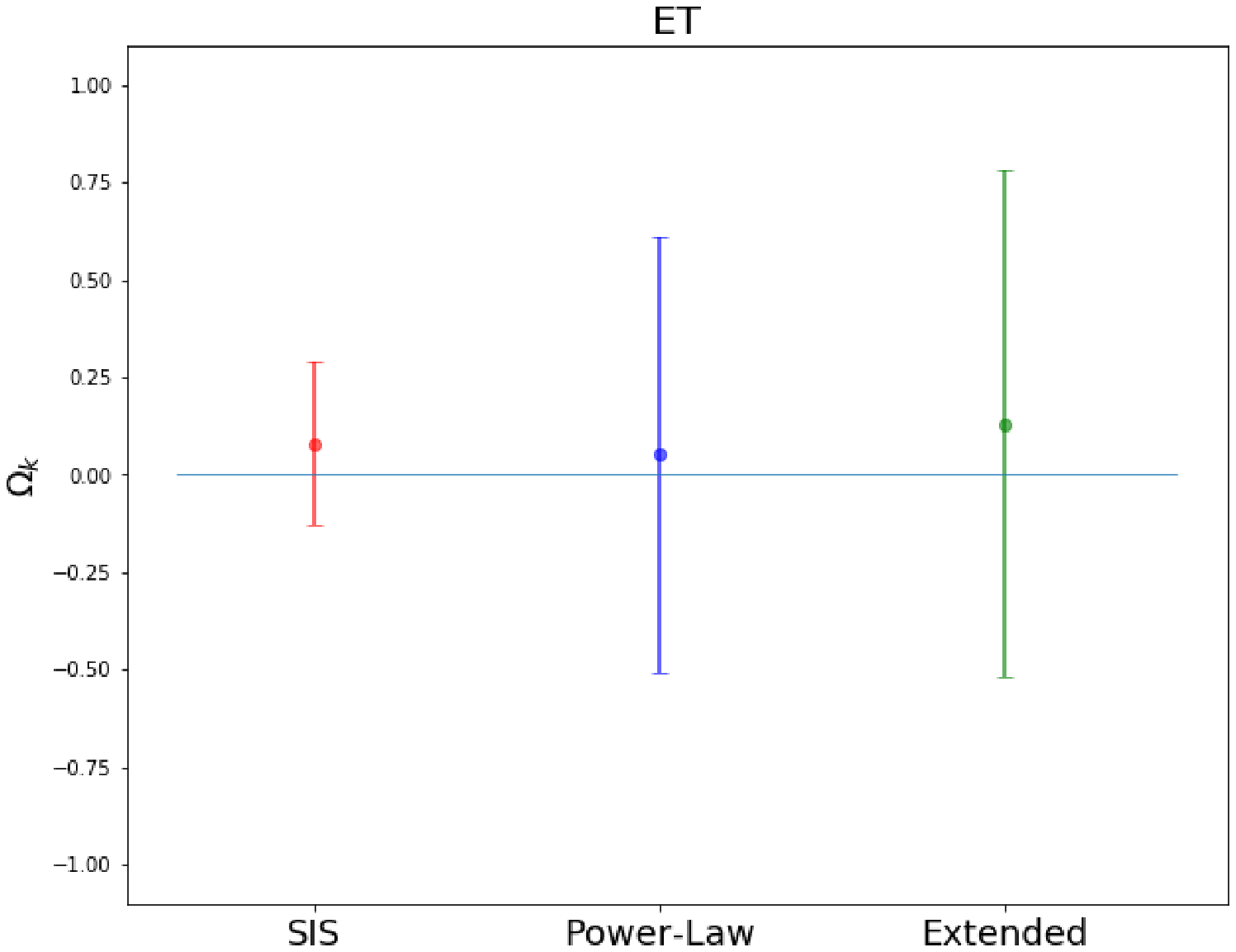}
\end{center}
\caption{Determination of cosmic curvature with different GW
detectors, future ground-based Einstein Telescope (ET) and satellite
GW observatory (DECIGO).}
\end{figure*}

Finally, in order to demonstrate the advantage of the
second-generation technology of space-borne GW detector, in Fig. 7
we compare the results of DECIGO with those obtained using the
third-generation ground-based Einstein Telescope (ET) \footnote{The
Einstein Telescope Project, https://www.et-gw.eu/et/}. See
\citet{Qi2019b} for detailed description of the simulation process
based on ET. It should be noted that adding the information brought
by such a space-based GW detector to the combination of LSST+DECIGO
does improve the $\Omega_K$ constraints significantly. More
specifically, the cosmic curvature is expected to be constrained
with an error smaller than $10^{-2}$, improving the sensitivity of
ET constraints by about a factor of 10 in the framework of three
kinds of spherically symmetric mass distributions (SIS, power-law
model, and extended power-law model) for the lensing galaxies. Such
conclusion could be not surprising, as although one would expect
that ET would be ten times more sensitive than current advanced
ground-based detectors, the neutron star-neutron star (NS-NS)
mergers and black hole-neutron star (BH-NS) mergers systems could
only be detected up to redshift $z\sim 2$ and $z\sim 5$,
respectively \citep{Cai17}. More importantly, benefit from the
fundamentally self-calibrating capability of space-based detectors,
the corresponding distance uncertainties for DECIGO may be reduced
to 1 percent accuracy at lower frequencies (in the frequency range
of 0.1-1 Hz) \citep{Cutler09}. This is to be compared with Fig.~1 in
\citet{Qi2019b}, whose forecast is for ET (with the specifications
foreseen at the time) the luminosity distance measurements could be
derived from 1000 observed GW events in the frequency range of
$1-10^4$ Hz. In summary, we do expect that the use of our technique,
i.e., using luminosity distance of standard sirens detected by the
second-generation technology of space-borne GW detector, would lead
to a stronger improvement in the direct measurement of the spatial
curvature in the early universe ($z\sim5.0$). However, in order to
investigate this further, the mass density profiles of early-type
galaxies should be properly taken into account \citep{Qi18}.

\section{Conclusions}

The spatial curvature of the Universe has been one of the most
important cosmological parameters in modern cosmology. Its value, or
even its sign would help us rule out the standard cosmological
paradigm or even point to the presence of new physics. In this work,
we have quantified the ability of DECIGO, a future Japanese space
gravitational-wave antenna in combination with galactic-scale strong
lensing systems expected to be detected by LSST, to improve the
current constraints on the cosmic curvature in the redshift range
$z\sim 5$. In the framework of the well-known distance sum rule
\citep{Rasanen2015}, the perfect redshift coverage of the standard
sirens observed by DECIGO, compared with lensing observations
including the source and lens from LSST, makes such
cosmological-model-independent test more natural and general. While
we exploited a commonly used Singular Isothermal Ellipsoid (SIE)
model to describe the mass distribution of lensing galaxies, we also
use Power-law model and Extended power-law model to better assess
the effect of lens model on measuring the cosmic curvature.

In the case of the simplest SIS lens model, due to the significant
increase of well-measured strong lensing systems and standard
sirens, one could expect the most stringent fits on the cosmic
curvature which has been improved by two orders of magnitudes
compared with the previous results obtained in \citet{Rasanen2015}.
Such precision is competitive with that derived from the Planck CMB
power spectra (TT, TE, EE+lowP) data \citep{Planck Collaboration}.
For the second lens model, we have considered the power-law mass
density profile in which the mass density power-law index of massive
elliptical galaxies evolves with redshift. Our findings indicate
that the constraint on the cosmic curvature in this case is still
very strong compared with that found for the SIS model. However, our
analysis reveals the strong degeneracy between the spatial curvature
and the redshift evolution of power-law lens index parameter.
Compared with the previous results focusing on a constant power-law
slope, we show that the estimation of $\Omega_K$ is more sensitive
to the measurement of $\gamma_1$, i.e., a significant redshift
evolution of the mass density power-law index will result in a
larger cosmic curvature. Therefore, additional observational
information, such as dedicated follow-up imaging of the lensed
images for a sample of individual lenses is necessary in this case.
Focusing on the performance of the extended power-law lens model, in
which the total density slope, luminosity density slope and the
anisotropy of stellar velocity dispersion are taken as free
parameters, the combined LSST+DECIGO data will improve the
constraints significantly with respect to current results, showing
the high constraining power that can be reached by the forthcoming
surveys. More interestingly, the extended power-law lens model is
more suitable to estimate the mass distribution of baryonic matter
and dark matter in the early-type galaxies. Specially, the addition
of DECIGO to the combination of LSST+DECIGO does improve the
constraint on the luminosity density slope and the anisotropy of the
stellar velocity dispersion significantly. In this case, our results
highlight the importance of investigating the luminosity density
slope and the anisotropy of the stellar velocity dispersion through
auxiliary data, especially integral field unit (IFU) spectroscopic
data in view of upcoming surveys.

What we are more concerned about is the advantage of higher quality
data sets from the second-generation technology of space-borne GW
detector, compared with the third-generation ground-based Einstein
Telescope (ET). For this purpose, our analysis demonstrates that the
cosmic curvature is expected to be constrained with an error smaller
than $10^{-2}$, improving the sensitivity of ET constraints by about
a factor of 10 in the framework of three kinds of lens models. In
summary, our paper highlights the benefits of synergies between
DECIGO and LSST in constraining the physical mechanism of cosmic
acceleration or new physics beyond the standard model, which could
manifest itself through accurate determination of the spatial
curvature of the Universe.

\vspace{0.5cm}

This work was supported by National Key R\&D Program of China No.
2017YFA0402600; the Ministry of Science and Technology National
Basic Science Program (Project 973) under Grants No. 2014CB845806;
the National Natural Science Foundation of China under Grants Nos.
12021003, 11690023, 11633001, and 11373014; Beijing Talents Fund of
Organization Department of Beijing Municipal Committee of the CPC;
the Strategic Priority Research Program of the Chinese Academy of
Sciences, Grant No. XDB23000000; and the Interdiscipline Research
Funds of Beijing Normal University. M.B. was supported by Foreign
Talent Introducing Project and Special Fund Support of Foreign
Knowledge Introducing Project in China. He is also grateful for
support from Polish Ministry of Science and Higher Education through
the grant DIR/WK/2018/12.

\end{document}